\documentclass[aps,pra,showpacs,superscriptaddress,12pt]{revtex4-1}

\usepackage{amsmath,amsfonts,amssymb,amsbsy}
\usepackage{graphicx}
\usepackage{color}
\usepackage[colorlinks=true,citecolor=red,linkcolor=blue]{hyperref}
\usepackage{enumerate}

\begin{document}

\title{Anisotropy Control in Photoelectron Spectra:\\ A Coherent Two-Pulse Interference Strategy}

\author{R. Chamakhi}
\affiliation{LSAMA Department of Physics, Faculty of Science of Tunis, University of Tunis El Manar, 2092 Tunis, Tunisia.}

\author{M. Telmini}
\affiliation{LSAMA Department of Physics, Faculty of Science of Tunis, University of Tunis El Manar, 2092 Tunis, Tunisia.}

\author{O. Atabek}
\affiliation{Institut des Sciences Mol\'eculaires d'Orsay (ISMO) UMR CNRS 8214, Univ. Paris-Sud, Universit\'e Paris-Saclay, F-91405 Orsay, France.}

\author{E. Charron}
\affiliation{Institut des Sciences Mol\'eculaires d'Orsay (ISMO) UMR CNRS 8214, Univ. Paris-Sud, Universit\'e Paris-Saclay, F-91405 Orsay, France.}
    
\date{\today}

\begin{abstract}
Coherence among rotational ion channels during photoionization is exploited to control the anisotropy of the resulting photoelectron angular distributions at specific photoelectron energies. The strategy refers to a robust and single parameter control using two ultra-short light pulses delayed in time. The first pulse prepares a superposition of a few ion rotational states, whereas the second pulse serves as a probe that gives access to a control of the molecular asymmetry parameter $\beta$ for individual rotational channels. This is achieved by tuning the time delay between the pulses leading to channel interferences that can be turned from constructive to destructive. The illustrative example is the ionization of the E($^1\Sigma_g^+$) state of Li$_2$. Quantum wave packet evolutions are conducted including both electronic and nuclear degrees of freedom to reach angle-resolved photoelectron spectra. A simple interference model based on coherent phase accumulation during the field-free dynamics between the two pulses is precisely exploited to control the photoelectron angular distributions from almost isotropic, to marked anisotropic.
\end{abstract}


\maketitle

%%%%%%%%%%%%%%%%%%%%%%
\section{Introduction}
%%%%%%%%%%%%%%%%%%%%%%

Quantum interference is one of the most reliable tools for controlling molecular dynamics both for electronic and nuclear degrees of freedom \cite{Shapiro_2003, JPB_50_234002, Nielsen_2000}. Laser-induced interference processes can be produced by a single pulse \cite{JPB_50_234002}, but also using two-pulse scenarii, as realized very often in coherent control strategies \cite{Shapiro_2003}. Time-resolved photoelectron angular distributions bring a very rich physical content based on an intricate combination of interference involving temporal and angular information whose control lead to challenging applications in electron-nuclear entanglement issues and photoelectron imaging, relevant in modeling biological processes \cite{Suzuki_2001}. At that respect, formal developments for calculating time-resolved photoelectron differential cross-sections owe much to the seminal work of T. Seideman and co-workers \cite{Seideman_rev}. More precisely, pure rotational \cite{Seideman_2000_1} and ro-vibrational motions \cite{Seideman_2000_2} have successively been introduced in complete treatments of angular distributions of diatomic systems, in terms of time and energy-dependent asymmetry parameter $\beta$. Even more challenging has been a full numerical implementation of such models to the study of non radiative transitions in pyrazine, a complex polyatomic molecule, by additionally including electronic relaxation, through the well-known ($S_2 \rightarrow S_1$) internal conversion mechanism \cite{Suzuki_2002, Suzuki_2003}. In these works, the dynamical description involves a short pump excitation, probed by a time-delayed ionization pulse and the role played (in alignment purpose for instance) of the pump field intensity, is appropriately accounted for by a non-perturbative treatment.  
Concurrently, time-resolved photoelectron spectra offer the possibility for quantum interference together with electron-nuclear entanglement through the ionization of rovibrational wave packets. Such schemes have been illustrated on the prototype alkali dimer Li$_2$ to which numerous experimental \cite{JCP_103_7269, PRA.66.043402, CPL.402.126} and theoretical \cite{JCP.114.1259, CPL.426.073301, CPB.19.033301, PRA.74.033407, JCP.147.144304} studies have been devoted. In particular, photoelectron kinetic energy distributions may be probed to reveal electron-nuclei entanglement resulting from the anisotropy of the diatomic molecule and transferring the rotational phase information in the photoelectron spectrum \cite{PRA.74.033407}. The control mechanism rests on an interference induced by a picosecond pulse (long enough duration as compared with the rotational period) involving different rotational channels of the ion. Searching for a systematic exploration of possible control achievements in photoelectron spectra, we have recently examined the role played by the pulse duration (or equivalently its bandwidth) from the pico- to femtosecond range, putting the emphasis on angular distributions rather than kinetic energies \cite{JCP.147.144304}. More precisely, it has been shown how the behavior of the asymmetry parameter $\beta$ is sensitive to the pulse duration and how this can be explained by a seemingly incoherent average over ion rotational channels. Still another two-color ($\omega$ and $2\omega$) laser induced electronic-vibrational coherence dynamics has very recently been successfully measured and interpreted in the case of the nitric oxide NO molecule \cite{Bandrauk_RC}. The interference in play corresponds to three-photon ($\omega+\omega+\omega$) and two-photon ($\omega+2\omega$) ionization processes creating a coherent superposition between two excited electronic states, and the control parameter is the relative phase of the pulses \cite{Yuan_2016}. This last work shows how the angular resolved photoelectron spectrum encodes the information of the time-dependent superposition molecular state and confirms the experimental feasibility of such control achievements.

The present work is devoted to a complete analysis of an interference scenario to provide control strategies based on specific constructive versus destructive schemes  of photoelectron angular distributions of Li$_2$ referring to a train of two ultra-short (femtosecond) linearly polarized light pulses, the control knob being merely the time delay $\tau$ between them. Full quantum calculations are interpreted using a simple analytical model describing the interference between accessible ion rotational channels, including the phase accumulated over $\tau$ during the evolution between the pulses. We ultimately show that this single parameter $\tau$ can be used to exert an efficient and robust control on the molecular asymmetry parameter $\beta$ varying it in a wide range, from values close to $\beta=2$, signature of an anisotropic distribution pointing along the polarization axis, to values close to $\beta=0$, typical for an isotropic distribution. We also show that this control can be used to disentangle the normally mixed contributions due to the different ion rotational states involved. As molecular one-photon ionization experiments with varying pulse duration is becoming an important issue in High Order Harmonic Generation\,\cite{Nature_460_972, Nature_Ph_9_93, Science_312_424, PRL.108.263003, JO.19.124016} and in large-scale free-electron laser (FEL) experiments\,\cite{FELs}, these results fall within the expanding field of ultra-fast photoelectron spectroscopy.

The paper is organized as follows: In Section \ref{sec:theory}, we briefly recall the salient features of the quantum model describing the Li$_2$ dimer excited by two-pulses acting on both electronic and nuclear dynamics. This section is also devoted to the development of the simplified analytical model which will later be used to reach a complete rationalization of the role played by ion rotational channels in the interference process affecting photoelectron energy and angular distributions. Section \ref{sec:results} presents the study of kinetic energy and angular distributions at specific energies, with a complete description of the coherent control strategy of the asymmetry parameter $\beta$. Conclusions and some new perspectives are finally presented in Section \ref{sec:conclusion}.

%%%%%%%%%%%%%%%%%%%%%%%%%%%%%%%%%%%%%%%
\section{The molecule-plus-field model}
\label{sec:theory}
%%%%%%%%%%%%%%%%%%%%%%%%%%%%%%%%%%%%%%%

%%%%%%%%%%%%%%%%%%%%%%%%%%%%%%%%%%%%%%%%%%%%%%
\subsection{The three-step ionization process}
%%%%%%%%%%%%%%%%%%%%%%%%%%%%%%%%%%%%%%%%%%%%%%

In this study, Li$_2$ photoionization proceeds through a three-step excitation \cite{JCP_103_7269, PRA.66.043402, CPL.402.126, PRA.74.033407}, namely: (i) A linearly polarized continuous wave (cw) resonant light field launches the initial ground rovibrational level ($v_X=0, N_X=0$) of the ground electronic state X($^1\Sigma_g^+$) on the first excited A($^1\Sigma_u^+$) electronic state, preparing the rovibrational level ($v_A=0, N_A=N_X+1=1$) ; (ii) A second cw laser resonantly prepares one of the two rovibrational levels ($v_E=0, N_E=N_A \pm 1=0,2$) of a second excited state E($^1\Sigma_g^+$). Note that the symmetries and Franck-Condon overlaps are favorable for such transitions, which were already observed\,\cite{JCP.108.9259, JCP.114.10311, PRA.68.043409, CPL.402.27}. In the following we assume well-separated dynamics from each of these rotational states, namely $N_E=0$ and $N_E=2$ ; (iii) The third step corresponds to the ionizing probe preparing a superposition of rotational channels in the continuum of the Li$_2^+$ ground electronic state X($^2\Sigma_g^+$). These channels are labeled by the rotational quantum number $N^+$ of the ion resulting from a photon absorption from the rotational state $N_E$ and from the ejection of a $p$-electron. Considering the fact that both a photon absorption and a $p$-electron escape modify the angular momentum by $\pm 1$, the ion rotational channels which are ultimately reached are $N^+=0,2$ for the initial state $N_E=0$, or $N^+=0,2,4$ for the initial state $N_E=2$. The corresponding rotational levels are separated by 6$B$ for $N^+=0,2$, and by 20$B$ for $N^+=0,2,4$, $B$ being the ion rotational constant for $v_+=0$, taken as $B \simeq 0.5$\,cm$^{-1}$\,\cite{JCP.147.144304}.

\begin{figure}[t]
	\centering
	\includegraphics[width = 0.99\columnwidth]{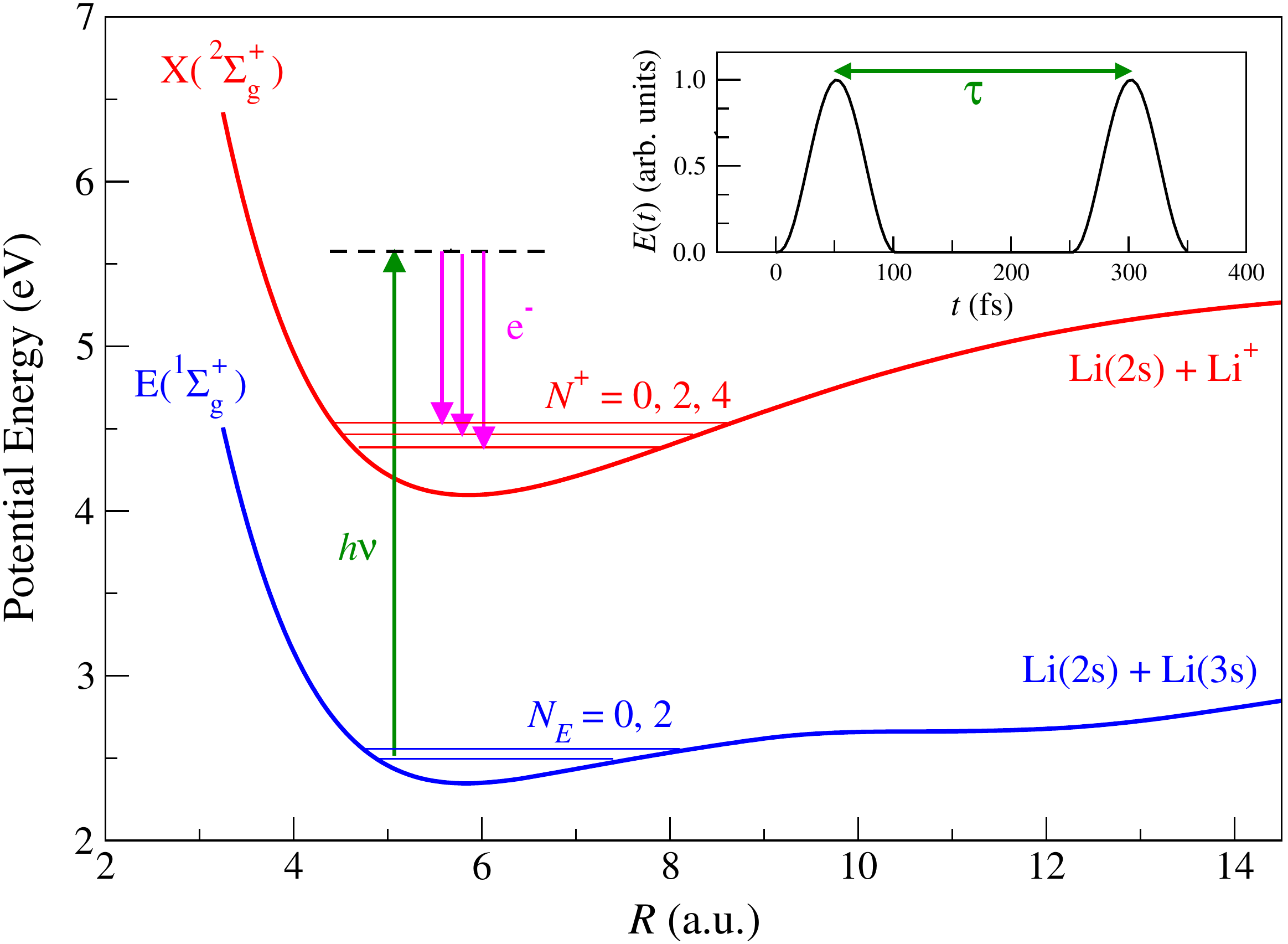}
	\caption{(Color online) Potential energy curves of the Li$_2$ E($^1\Sigma_g^+$) and Li$_2^+$ X($^2\Sigma_g^+$) electronic states describing the third step of the excitation process, in blue and red solid lines respectively. The corresponding rotational levels are indicated by horizontal lines. The inset illustrates the envelope of the two ultra-short pulses delayed by $\tau$.}
	\label{fig:model}
\end{figure}

The total energy brought in the system by the probe pulse is shared among the ion rovibrational energy and the photoelectron kinetic energy. Fig.~\ref{fig:model} displays the potential energy curves of the E(Li$_2$) and X(Li$_2^+$) electronic states together with the  rotational levels in consideration. Due to similar values of the equilibrium position and vibrational frequency of these states, a Franck-Condon vertical launching of the initial vibrationless wave function does not change the corresponding vibrational levels; \emph{i.e.}, $v_X=v_A=v_E=v_+=0$. The probe field illustrated in the inset of Fig.~\ref{fig:model} is built from two ultra-short identical sine-square pulses, both of full width at half maximum (FWHM) $T=50$\,fs, delayed by a tunable time $\tau$, typically in the femtosecond to picosecond range. More precisely, the pulsed electric field is written as
\begin{equation}
\mathbf{E}(t) = E(t)\,\mathbf{\hat{e}} = E_0\,\Big[f(t)+f(t-\tau)\Big] \cos(\omega t)\,\mathbf{\hat{e}}\,,
\label{eq:field}
\end{equation}
with the individual pulse envelop
\begin{equation}
f(t) = \sin^2\!\left(\frac{\pi t}{2T}\right)
\quad
\forall t \in [0,2T]\,,
\label{eq:field_shape}
\end{equation}
where $f(t)=0$ for $t \leqslant 0$ and $ t \geqslant 2T$. The pulses are linearly polarized along the unit vector $\mathbf{\hat{e}}$, with a peak amplitude $E_0$ weak enough to remain in the linear excitation regime. It is worth noting that such ultra-short pulses have a bandwidth $\Delta \omega$ large enough to encompass all accessible rotational levels $N^+$, since $\Delta \omega \gg 20B$. They can thus induce interfering ionization paths among the corresponding channels. In the following, the space-fixed $z$-axis is taken to be along the field polarization vector.

Hereafter we focus on the last step of the excitation process leading to two separated ionization routes starting either from $N_E=0$ or $N_E=2$, taken independently. The possible interference mechanism rests on the phase accumulation, due to the time delay $\tau$, affecting the different ion rotational channels $N^+$.

%%%%%%%%%%%%%%%%%%%%%%%%%%%%%%%%%%%%%%%%%%%%%%%%%%%%%
\subsection{Multichannel close-coupled equations formalism}
%%%%%%%%%%%%%%%%%%%%%%%%%%%%%%%%%%%%%%%%%%%%%%%%%%%%%

The time evolution of the wave packet  takes place on the two electronic states,  E($^1\Sigma_g^+$) of Li$_2$ (involving a component $\Phi_E$) and  X($^2\Sigma_g^+$) of \{Li$_2^+ + 1\mathrm{e}^-$\} (with a component $\Phi_+$), written as
\begin{equation}
\Psi(\mathbf{r}_\mathrm{e},\mathbf{R};t)=\Phi_E(\mathbf{r}_\mathrm{e},\mathbf{R};t) + \Phi_+(\mathbf{r}_\mathrm{e},\mathbf{R};t)\,,
\label{eq:wavepacket}
\end{equation}
where $\mathbf{r}_\mathrm{e}=\{\mathbf{r}_\mathrm{c},\mathbf{r}\}$ collectively stands for the electronic coordinates, and $\mathbf{R}$ for the nuclear ones. The escaping electron is described by its coordinate $\mathbf{r}$, the remaining core electrons being accounted for by $\mathbf{r}_\mathrm{c}$. Each component $\Phi_E$ and $\Phi_+$ is given in terms of a Born-Oppenheimer (BO) product
\begin{subequations}
\begin{eqnarray}
\label{eq:componentE}
\Phi_E(\mathbf{r}_\mathrm{e},\mathbf{R};t) & = & \chi_E(\mathbf{R};t)\, \tilde{\psi}_E(\mathbf{r}_\mathrm{e};R)\\
\Phi_+(\mathbf{r}_\mathrm{e},\mathbf{R};t) & = & \chi_+(\mathbf{r},\mathbf{R};t)\, \tilde{\psi}_X(\mathbf{r}_\mathrm{c};R)
\label{eq:componentX}
\end{eqnarray}
\end{subequations}
of the time-independent electronic wavefunctions $\tilde{\psi}_E(\mathbf{r}_\mathrm{e};R)$ or $\tilde{\psi}_X(\mathbf{r}_\mathrm{c};R)$ resulting from a fixed internuclear distance scheme, with time-dependent wave packets $\chi_E(\mathbf{R};t)$ or $\chi_+(\mathbf{r},\mathbf{R};t)$ to be further calculated.

As usual, the electronic wavefunction $\tilde{\psi}_E$ is expressed in the molecular frame with an angular part for the $3s\sigma$ Rydberg electron described by the ground state spherical harmonic
\begin{equation}
\tilde{\psi}_E(\mathbf{r}_\mathrm{e};R) = \psi_E(\mathbf{r}_\mathrm{c},r;R)\,Y_{00}(\mathbf{\hat{r}})\,.
\end{equation}
As for $\tilde{\psi}_X(\mathbf{r}_\mathrm{c};R)$ in Eq.~(\ref{eq:componentX}), it is nothing but the BO electronic wavefunction of the Li$_2^+$ (X) state depending solely on core electrons, whereas the escaping $p$-electron dynamics (coordinate $\mathbf{r}$, angular momentum $\ell=1$) is accounted for by $\chi_+(\mathbf{r},\mathbf{R};t)$.
Moreover, the escaping electron is appropriately described in the laboratory frame, leading to an expansion on the time-independent electronic continuum basis
\begin{equation}
\psi_{\ell m}(\mathbf{r},\varepsilon;R)=\psi_{\ell}(r,\varepsilon;R)Y_{\ell m}(\mathbf{\hat{r}})\,,
\end{equation}
which reads as
\begin{equation}
\chi_+(\mathbf{r}, \mathbf{R}; t) = \sum_{m} \int \!d\varepsilon\;\chi_{\ell m}^+(\mathbf{R},\varepsilon;t)\,\psi_{\ell m}(\mathbf{r},\varepsilon;R)\,,
\label{eq:chi+}
\end{equation} 
with time-dependent nuclear wave packets $\chi_{\ell m}^+(\mathbf{R},\varepsilon;t)$ to be calculated. In the following the electron angular momentum $\ell$ is fixed to $\ell = 1$ because of the emission of a $p$-type electron, and we could therefore omit this index but we decided to keep it in order to conserve a general formalism. The summation in Eq.\,(\ref{eq:chi+}) is actually limited to $m=-1,0,+1$, projections of $\ell$ in the laboratory frame. $\varepsilon$ is the escaping electron kinetic energy.

Finally, the angular parts of the nuclear wave packets are expressed in terms of Wigner rotation matrices $\mathcal{D}_{M\Lambda}^N(\mathbf{\hat{R}})$ \cite{Zare_1988}
\begin{subequations}
\begin{eqnarray}
\chi_{E}(\mathbf{R}; t) & = & \chi_M^{N_E}(R; t)\, {\mathcal{D}_{M,0}^{N_E^{\;\;*}}}(\mathbf{\hat{R}})\,,\\
\chi_{\ell m}^+(\mathbf{R}, \varepsilon; t) & = & \sum_{N^+} \chi_{M^+}^{N^+}(R,\varepsilon;t)\,{\mathcal{D}_{M^+,0}^{ N^{+\,*}}}(\mathbf{\hat{R}})\,,
\end{eqnarray}
\label{eq:chiEchiM+}
\end{subequations}
where $M$ and $M^+$ are the projections of the molecular rotational angular momenta of Li$_2$ and Li$_2^+$ on the polarization axis with
\begin{equation}
M^+=M-m\,.
\end{equation}
Note that in our study we start from $N_X=0$ and therefore from $M_X=0$. In addition, $M$ is conserved during the resonant multiphoton excitation process proposed here. In Eq.\,(\ref{eq:chiEchiM+}) $M$ is therefore limited to the value $M=0$ and $M^+$ can take the values $M^+=-m=+1,0,-1$.

The time-dependent Schr\"{o}dinger equation
\begin{equation}
i\hbar\partial_t \Psi(\mathbf{r}_\mathrm{e},\mathbf{R};t) =
\hat{\mathcal{H}}\,\Psi(\mathbf{r}_\mathrm{e},\mathbf{R};t)
\end{equation}
is solved by introducing the expansion of Eq.\,(\ref{eq:wavepacket}) and projecting on the electronic and rotational basis sets, which results in the following set of close-coupled equations written in the length gauge as
\begin{subequations}
\begin{eqnarray}
i\hbar\partial_t\,\chi_M^{N_E} & = &
\mathcal{H}_{N_E}^{E}\chi_M^{N_E}
-E(t)\!\!\sum_{N^+,m}\!\mathcal{W}_{m}^{N^+}\chi_{M^+}^{N^+}\\
i\hbar\partial_t\,\chi_{M^+}^{N^+} & = &
\Big(\mathcal{H}_{N^+}^{X}+\varepsilon\Big)\,\chi_{M^+}^{N^+}
-E(t)\,\mathcal{W}_{m}^{N^+*}\,\chi_{M}^{N_E}\;\;\;\;
\end{eqnarray}
\label{eq:ccequations}
\end{subequations}
The nuclear Hamiltonians are given as
\begin{equation}
\mathcal{H}_{N}^{\alpha} =
-\frac{\hbar^2}{2\mu}\,\Big[
  \frac {\partial^2}{\partial R^2}
- \frac{N(N+1)}{R^2}
\Big] + V_\alpha(R)\,,
\label{eq:Hamiltonians}
\end{equation}
with $\alpha = E$ (for Li) or $X$ (for Li$^+$). The evaluation of the coupling terms $\mathcal{W}_{m}^{N^+}$ requires a unitary molecule-to-laboratory frame transformation involving short-range quantum defects $\mu_{\Lambda=0,1}$, together with H\"{o}nl-London rotational factors \cite{PRA.2.353, PRA.6.185, JCP.66.5584, JCP.73.3338, JCP.74.3388}. A full derivation has been given in our previous work \cite{PRA.74.033407} resulting in
\begin{equation}
\mathcal{W}_{m}^{N^+} = \mathcal{C}_{m}^{N^+}\,d_{\ell}(\varepsilon,R)\,,
\end{equation}
where $\mathcal{C}_{m}^{N^+}$ is a proportionality factor build in terms of Wigner 3-j coefficients \cite{Zare_1988} and the above mentioned frame transformation matrix elements. It is crucial to take into account the rotational dynamics for angle resolved photoelectron distributions, and this is introduced through Wigner matrices in the basis set representation of the wave functions in Eqs.\,(8a) and (8b). The initial state chosen does not assume any pre-alignment of the molecule. All degrees of freedom are thus considered through basis set expansions, except the internuclear distance R which is dynamically treated through close-coupled equations (11a) and (11b). As for the, in principle energy and $R$-dependent ionization dipole $d_{\ell}(\varepsilon,R)$, it is given by the integral
\begin{equation}
d_{\ell}(\varepsilon,R) =
\Big\langle
\psi_{\ell}\,\psi_X
\,\Big|\, r \,\Big|\,
\psi_E
\Big\rangle_{r,\mathbf{r}_\mathrm{c}}.
\label{eq:dpole}
\end{equation}
Restricting to some limited range of photoelectron energies and internuclear distances, a Condon-type approximation is used hereafter to fix $d(\varepsilon,R)$ merely as a constant.

\subsection{Photoelectron spectra and interference patterns}

The time propagation of the nuclear wave packets $\chi_M^{N_E}(R;t)$ and $\chi_{M^+}^ {N^+}(R,\varepsilon;t)$ is achieved using a third-order split operator technique \cite{JCompP.47.412, JCP.108.3922} together with initial conditions at $t=0$ involving a single ro-vibrational state for $\chi_M^{N_E}(R;0)$ (with $v_E=0,N_E=0$ or 2 and $M=0$), and $\chi_{M^+}^{N^+}(R,\varepsilon;0)=0$. The initial ro-vibrational state is calculated using the Numerov algorithm \cite{Numerov}. The integration of the coupled time-dependent Schrödinger equations (11a) and (11b) is performed following Refs. \cite{PRA.74.033407} and \cite{JCP.108.3922}, using the split-operator method in the rotating-wave approximation with the time step $\delta t = 4$\,fs. Photoelectron angular distributions are obtained by projecting $\chi_+(\mathbf{r}, \mathbf{R}; t_t)$ at the final time $t_f=2T+\tau$ corresponding to the switch-off of the two-pulse sequence, on energy-normalized out-going waves, in the $\mathbf{k} \equiv (k,\mathbf{\hat{k}})$ direction. More precisely these states are written in terms of their angular part expanded on spherical harmonics as
\begin{equation}
\varPhi_\ell(\mathbf{k},\mathbf{r};R) = i e^{-i\xi}
\sum_m Y_{\ell m}^*(\mathbf{\hat{k}})\,\psi_{\ell m}(\mathbf{r},\varepsilon;R)\,,
\label{eq:Phi}
\end{equation}
where the momentum $k$ is related to the asymptotic electron kinetic energy $\varepsilon$ by $\hbar^2k^2=2m\varepsilon$. Note also that for the exit channel $N^+$ energy conservation yields the following relation: $\varepsilon_{N^+} = E_{N_E} + \hbar\omega - E_{N^+}$ between the photoelectron energy $\varepsilon_{N^+}$, the energy of the initial rovibrational state $E_{N_E}$, the photon energy $\hbar\omega$ and the energy of the final rovibrational state $E_{N^+}$. This relation is strictly valid in the cw regime. In Eq.\,(\ref{eq:Phi}) $\xi$ denotes the Coulomb phase-shift of the $\ell=1$ outgoing wave. As a consequence, the angle and energy-resolved photoelectron distribution is given by \cite{PRA.74.033407}
\begin{equation}
P(\varepsilon,\mathbf{\hat{k}}) = \int d\mathbf{R}\,\left| \int \varPhi^*_\ell(\mathbf{k},\mathbf{r};R)\,\chi_+(\mathbf{r},\mathbf{R}; t_f)\,d\mathbf{r} \right|^2.
\end{equation}
This last equation can be further simplified using orthogonality rules, resulting ultimately in \cite{PRA.74.033407}
\begin{equation}
P(\varepsilon,\mathbf{\hat{k}}) \propto \sum_{N^+,m}
\Big| Y_{\ell m}(\mathbf{\hat{k}}) \Big|^2
\int \!dR\,\Big|\,\chi_{M^+}^{N^+}(R,\varepsilon;t_f)\,\Big|^2\,.
\label{eq:angular}
\end{equation}
As for the total photoelectron spectrum we have to sum over the electron ejection angle $\mathbf{\hat{k}}=(\theta_k, \phi_k)$, giving
\begin{equation}
P(\varepsilon) \propto \sum_{N^+,m}
\,\int\!dR\,\Big|\,\chi_{M^+}^{N^+}(R,\varepsilon;t_f)\,\Big|^2\,.
\label{eq:totaldistribution}
\end{equation}
It has already been pointed out that the seemingly incoherent sums in Eq.~(\ref{eq:totaldistribution}) may still reveal interference effects when an exit rotational channel $N^+$ is reached from different initial states $N_E$ \cite{PRA.74.033407}. Pursuing the objective of clearly depicting the interference mechanism which is the salient point of this study, we wish to disentangle different coherent processes. This is why we are addressing separately a single initial level $N_E=0$ or $N_E=2$ instead of a coherent superposition of several ro-vibrational levels $N_E$, which remains a realistic choice when considering an initial preparation achievable by resonant cw excitation. The interference scheme we are hereafter referring to has therefore a different origin: a phase accumulation during the field-free evolution between the two ultra-short pulses depicted in the inset of Fig.~\ref{fig:model}.

More precisely, if we define the ionization amplitude $\mathcal{A}_{N^+}(\varepsilon)$ corresponding to the ionization by a single pulse, from a given initial state $N_E$ to the ionized state $N^+$ with a photoelectron energy $\varepsilon$, the photoelectron spectrum obtained with two delayed identical pulses is nothing but a coherent superposition of this amplitude and the same amplitude from the second pulse with a phase $\varphi(\varepsilon,\tau) = (\varepsilon-\varepsilon_{N^+})\tau/\hbar$ accumulated during $\tau$, given as
\begin{subequations}
\begin{eqnarray}
\label{eq:pcos17a}
P(\varepsilon,\tau) & = &
\left| \mathcal{A}_{N^+}(\varepsilon) +
\mathcal{A}_{N^+}(\varepsilon)\,e^{-i\varphi(\varepsilon,\tau)} \right|^2\;\;\\
\label{eq:pcos17b}
& = & 4\big|\mathcal{A}_{N^+}(\varepsilon)\big|^2
\cos^2\left[\frac{(\varepsilon-\varepsilon_{N^+})\tau}{2\hbar}\right].
\end{eqnarray}
\label{eq:pcos}
\end{subequations}
It thus appears that $P(\varepsilon,\tau)$ presents an oscillatory behavior as a function of both the electron kinetic energy $\varepsilon$ and the delay $\tau$ between the two pulses. When this time is taken as a control knob, a constructive interference is therefore expected at a specific energy $\bar{\varepsilon}$ for the delay
\begin{equation}
\tau^{\mathrm{c}}_n = 2n\,\tau^*_{N^+}(\bar{\varepsilon})\;\;\mathrm{with}\;\;n=0, 1, 2...,
\label{eq:Tcons}
\end{equation}
where
\begin{equation}
\tau^*_{N^+}(\bar{\varepsilon}) = \frac{\pi\hbar}{|\bar{\varepsilon}-\varepsilon_{N^+}|}\,.
\end{equation}
For this delay $\tau^{\mathrm{c}}_n$ the photoelectron spectrum is given by
\begin{equation}
P\big(\varepsilon,\tau^{\mathrm{c}}_n\big) =
4\big|\mathcal{A}_{N^+}(\varepsilon)\big|^2
\cos^2\!\Big[n\pi\Big(\frac{\varepsilon-\varepsilon_{N^+}}{\bar{\varepsilon}-\varepsilon_{N^+}}\Big)\Big].
\label{eq:pcos_cons}
\end{equation}
It oscillates as a function of the electron kinetic energy $\varepsilon$ and it reaches indeed a maximum of $4|\mathcal{A}(\bar{\varepsilon})|^2$ for the particular kinetic energy $\varepsilon=\bar{\varepsilon}$.

Symmetrically, a destructive interference is expected at the energy $\bar{\varepsilon}$ for the delay
\begin{equation}
\tau^{\mathrm{d}}_n = (2n+1)\,\tau^*_{N^+}(\bar{\varepsilon})\;\;\mathrm{with}\;\;n=0, 1, 2...
\label{eq:Tdes}
\end{equation}
yielding
\begin{equation}
P\big(\varepsilon,\tau^{\mathrm{d}}_n\big) =
4\big|\mathcal{A}_{N^+}(\varepsilon)\big|^2
\cos^2\!\Big[(2n+1)\frac{\pi}{2}\Big(\frac{\varepsilon-\varepsilon_{N^+}}{\bar{\varepsilon}-\varepsilon_{N^+}}\Big)\Big]
\label{eq:pcos_des}
\end{equation}
For this time delay $\tau^{\mathrm{d}}_n$ the photoelectron spectrum is still expected to oscillate with $\varepsilon$, but it reaches a minimum at the energy $\varepsilon=\bar{\varepsilon}$, with
$P\big(\bar{\varepsilon},\tau^{\mathrm{d}}_n\big)=0$.

The ionization amplitude $\mathcal{A}_{N^+}(\varepsilon)$ seen in Eqs.\,(\ref{eq:pcos}), (\ref{eq:pcos_cons}) and (\ref{eq:pcos_des}) is, in principle, specific to each exit channel $N^+$ and according to Eq.\,(\ref{eq:totaldistribution}) the total photoelectron spectrum is given by an incoherent sum over all ion rotational levels $N^+$. One may expect that such an incoherent sum could wash out the regular oscillation patterns predicted in Eqs.\,(\ref{eq:pcos_cons}) and (\ref{eq:pcos_des}) by the interference model that we have developed previously for a single exit channel. However, for a pulse duration $T$ much shorter than the rotational period $\pi\hbar/B \simeq 33$\,ps one can consider that the different photoelectron energies $\varepsilon_{N^+}$ associated with these exit channels are almost degenerate. In such a case we expect that the interference model developed previously will still hold. For the implementation of this model we will hereafter simply use Eqs.\,(\ref{eq:pcos}), (\ref{eq:pcos_cons}) and (\ref{eq:pcos_des}), replacing $\varepsilon_{N^+}$ with the particular value obtained for $N^+=N_E$ and replacing the envelop $|\mathcal{A}_{N^+}(\varepsilon)|^2$ with the photoelectron spectrum $P(\varepsilon)$ calculated for a \emph{single} pulse. In summary, it is worthwhile mentioning that we are considering two models:
\begin{enumerate}[(i)]
    \item the first one dealing with the full quantum description of the ionization dynamics from $t=0$ to $t_f$, as given by Eq.\,(\ref{eq:totaldistribution});
    \item the second one consisting in the interference model we have developed using an inter-pulse phase accumulation. This last model makes use, through Eq.\,(\ref{eq:pcos17b}), of the envelop $|\mathcal{A}_{N^+}(\varepsilon)|^2$ calculated with the first model (i) using a single pulse.
\end{enumerate}

%%%%%%%%%%%%%%%%%%%%%%%%%%%%%%%%%%%%%%%%%%%%%%%%%%%%%%
\section{Results: From interpretation to control}
\label{sec:results}
%%%%%%%%%%%%%%%%%%%%%%%%%%%%%%%%%%%%%%%%%%%%%%%%%%%%%%

This section separately addresses the interpretation and control of photoelectron energy and angular distributions. The electronic potential energy curves are taken from Ref.\,\cite{CP.92.263}. The reduced mass is $\mu=3.4695$\,Da. In weak fields the ionization dipole of Eq.\,(\ref{eq:dpole}), together with the electric field amplitude enter as simple scaling factors and their values need not be specifically defined. The pulse duration $T$ and the rotational constant (which is very close for the $E$ and $X$ electronic states) have already been defined, whereas the delay $\tau$ is considered as a tunable control parameter.

\begin{figure}[t]
	\centering
	\includegraphics[width = 0.99\columnwidth]{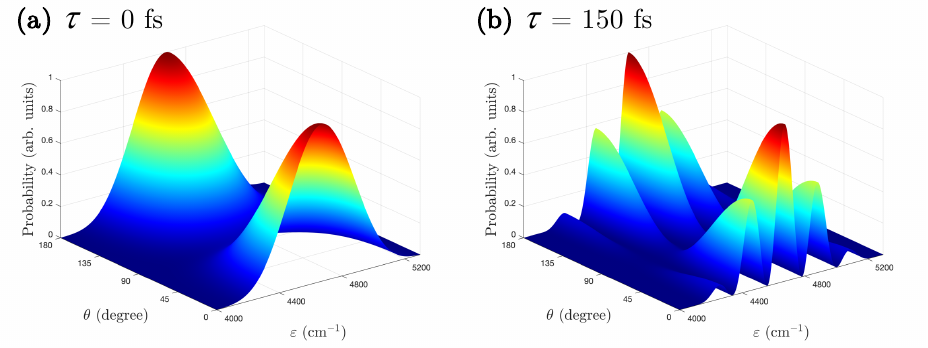}
	\caption{(Color online) Angle ($\theta$) and energy ($\epsilon$) resolved normalized photoelectron spectra as obtained from the full quantum model of Eq.\,(\ref{eq:angular}), for the initial rotational state $N_E=0$ when referring to either a single pulse (case labeled  $\tau=0$), on the left panel or two-pulses delayed by $\tau=150$\,fs, on the right panel.}
	\label{fig:angle_resolved_spectra}
\end{figure}

Starting from the initial level $N_E=0$, the photoelectron energies associated with the ion rotational levels are: $\varepsilon_0=4671.6$\,cm$^{-1}$ for $N^+=0$ and $\varepsilon_2=4668.6$\,cm$^{-1}$ for $N^+=2$. Starting from the initial level \mbox{$N_E=2$} yields the following energies: $\varepsilon_0=4674.6$\,cm$^{-1}$ for $N^+=0$, $\varepsilon_2=4671.6$\,cm$^{-1}$ for $N^+=2$ and $\varepsilon_4=4664.6$\,cm$^{-1}$ for $N^+=4$.

\begin{figure}[t]
\centering
\includegraphics[width = 0.99\columnwidth]{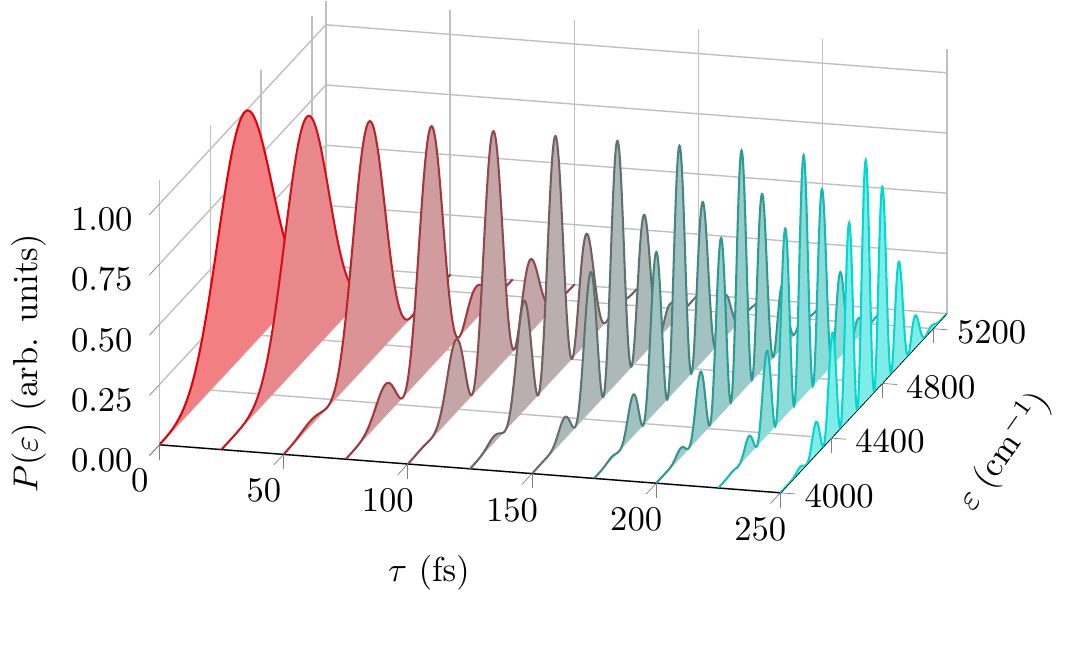}
\caption{(Color online) 3D view of the normalized photoelectron energy distributions as obtained from the full quantum model of Eq.\,(\ref{eq:totaldistribution}), for the initial rotational state $N_E=0$ and for a collection of inter-pulse time delays $\tau$.}
\label{fig:fullspectra}
\end{figure}

%%%%%%%%%%%%%%%%%%%%%%%%%%%%%%%%%%%%%%%%%%%%%%%%%%%%%%%%
\subsection{Photoelectron kinetic energy distributions}
%%%%%%%%%%%%%%%%%%%%%%%%%%%%%%%%%%%%%%%%%%%%%%%%%%%%%%%%

The angle-resolved photoelectron spectra resulting from the full time evolution model (i) based on Eq.\,(\ref{eq:angular}) for the initial rotational state $N_E=0$, when referring to a single pulse or two-pulses with a typical delay of $\tau=$150\,fs, are illustrated in Fig.~\ref{fig:angle_resolved_spectra}. One observes a $\theta$-dependence well peaked at $\theta =0$ or 180 degrees, symmetrical with respect to 90 degrees. The oscillating pattern obtained for $\tau=150$\,fs can be analyzed and interpreted by appropriately summing over angles $\theta$ (following Eq.\,(\ref{eq:totaldistribution})) to end up in total photoelectron spectra, as the ones illustrated in Fig.~\ref{fig:fullspectra} for a collection of inter-pulse time delays $\tau$. When $\tau$ is zero, the resulting single ultra-short pulse has a bandwidth of the order of $2\pi/T \simeq 660$\,cm$^{-1}$, much larger than the $N^+=0,2$ rotational levels separation which amounts to $6B \simeq 3$\,cm$^{-1}$. The corresponding photoelectron spectrum therefore presents a single broad peak with a maximum positioned around $\varepsilon_0 = 4671.6$\,cm$^{-1}$. Actually, the anisotropy of the $E$ and $X$ electronic states being relatively small \cite{PRA.74.033407}, the exit channel $N^+=0$ is favored in the branching ratio between the $N^+=0$ and $N^+=2$ ionization channels. In addition, increasing $\tau$ gives rise to an oscillatory interference patterns, as predicted by the interference model derived in Eq.\,(\ref{eq:pcos17b}).

For the ultimate goal of controlling the ionization dynamics, we now want to decipher this interference scheme. In particular, for the initial state $N_E=0$, the probability for a photoelectron to be emitted with the asymptotic energy $\bar{\varepsilon}_0=\varepsilon_0+\Delta$ is proportional to $\cos^2[\Delta\tau/(2\hbar)]$. The time delay $\tau$ can thus be chosen in such a way to reduce to zero this probability, producing a fully destructive interference. The appropriate choice would be $\tau = \pi\hbar/\Delta$. A completely similar analysis can be conducted for the initial state $N_E=2$ at the energy $\bar{\varepsilon}_2 = \varepsilon_2 + \Delta$. This would lead to the same delay for a destructive interference since $\tau^*_0(\bar{\varepsilon}_0)=\tau^*_2(\bar{\varepsilon}_2)$.

\begin{figure}[t]
\centering
\includegraphics[width = 0.99\columnwidth]{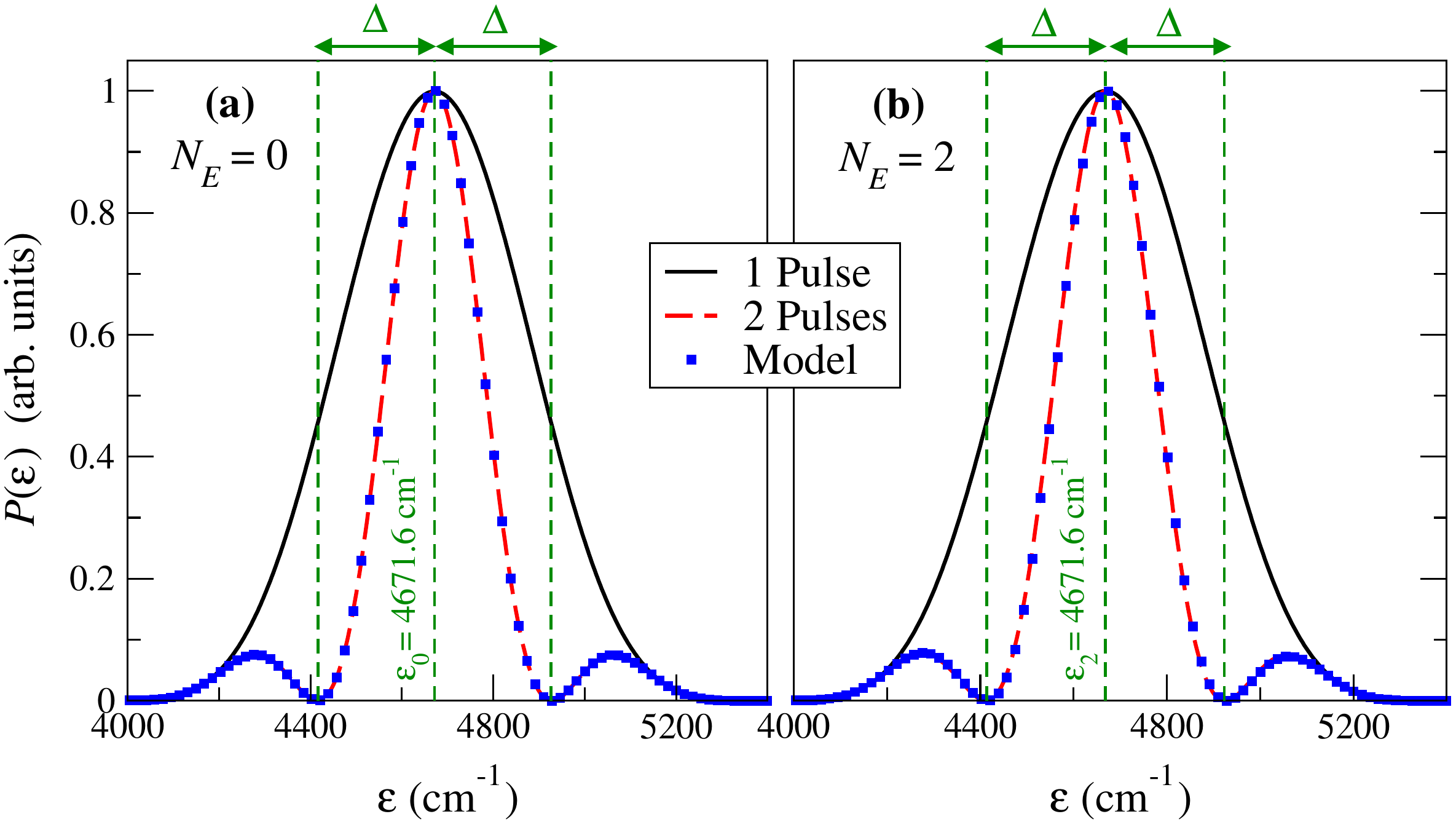}
\caption{(Color online) Photoelectron energy distributions for the two initial states $N_E=0$ panel (a) and $N_E=2$ panel (b). The solid black line is for the single pulse result. The red dashed line is for the full quantum calculation (i) with two pulses, whereas the blue squares are for the simplified interference model of Eq.\,(\ref{eq:pcos17b}). The inter-pulse delay chosen, $\tau=65$\,fs, produces a fringe separation of $2\Delta = 509.2$\,cm$^{-1}$.}
\label{fig:interpretation}
\end{figure}

We need of course to check the relevance and accuracy of this model as compared with an exact calculation. The results are displayed in Fig.~\ref{fig:interpretation}. The single pulse spectra corresponding to $\tau=0$ are indicated in solid black lines and are very broad compared to the rotational spacing. Panels (a) and (b) illustrate, for the initial states $N_E=0$ and $N_E=2$ respectively, the case of destructive interferences as obtained with $\Delta = 254.6$\,cm$^{-1}$ leading to $\tau=65$\,fs. The full quantum calculation of model (i) is displayed as a dashed red line, whereas the interference model of Eq.\,(\ref{eq:pcos17b}) is illustrated by blue square symbols. As can be judged by Fig.~\ref{fig:interpretation}, an excellent agreement is obtained between these two simulations, validating the accuracy of the interference model. In panel (a) the destructive interference totally suppressing the ionization signal at the energies $\varepsilon_0 \pm \Delta$, leads to a main peak positioned at $\varepsilon_0 = 4671.6$\,cm$^{-1}$, together with two satellites originating from the oscillations of the cosine square function. Panel (b) illustrates the symmetric situation for $N_E=2$. The main peak is at $\varepsilon_2 = 4671.6$\,cm$^{-1}$. It corresponds to a constructive interference for the $N^+=2$ exit channel and to a destructive interference at the photoelectron energies $\varepsilon_2 \pm \Delta$.

In summary, by appropriately tuning the time delay between the two ultra-short pulses, we can accurately alter the photoionization dynamics through an efficient interference effect seen in kinetic energy distributions. The choice of the control parameter, namely the inter-pulse delay $\tau$, is provided through a simple reduced interference model. This model turns out to be robust and accurate when checked against the exact calculation.

%%%%%%%%%%%%%%%%%%%%%%%%%%%%%%%%%%%%%%%%%%%%%%%%%%%%%%
\subsection{Photoelectron angular distributions}
%%%%%%%%%%%%%%%%%%%%%%%%%%%%%%%%%%%%%%%%%%%%%%%%%%%%%%

A more refined analysis of the photoionization can be carried out by examining photoelectron angular distributions as given by Eq.\,(\ref{eq:angular}) prior to the integration over electron ejection angles. Such distributions can be examined either integrated over all electron kinetic energies, or at specifically chosen energies. It is worth noting that, with linear polarization the only direction-dependence of $P(\varepsilon, \hat{k})$ is through  the azimuth angle $\theta$. From now on we will thus refer to $P(\varepsilon, \theta)$ instead of $P(\varepsilon, \hat{k})$ and the study will be conducted, as usually done, by referring to a single asymmetry parameter $\beta$ defined from the angular differential photoionization cross-section
\begin{equation}
\frac{d\sigma}{d\Omega} = \frac{\sigma}{4\pi}\,\Big[1 + \beta P_2(\cos\theta)\Big]\,,
\label{eq:crossection}
\end{equation}
where $\sigma$ is the total ionization cross-section over the full solid angle $\Omega = 4 \pi$. $P_2(x)$ denotes the second order Legendre polynomial with $P_2(x)=(3x^2-1)/2$. Note that Eq.\,(\ref{eq:crossection}) assumes a weak field linear response: it only takes into account single photon ionization processes. $\beta$ can be obtained unambiguously by recasting the photoelectron  angular distributions as
\begin{equation}
P(\varepsilon, \theta)=A(\varepsilon)+B(\varepsilon)\cos^2\theta\,,
\end{equation}
and by identifying the corresponding terms in $\cos\theta$ of Eq.\,(\ref{eq:crossection}). One finally gets the asymmetry parameter
\begin{equation}
\beta(\varepsilon) = \frac{2B(\varepsilon)}{3A(\varepsilon)+B(\varepsilon)}
\end{equation}
which characterizes the overall shape of photoelectron angular distributions. More precisely, peaked anisotropic distributions along or perpendicular to the polarization direction $\mathbf{\hat{e}}$ correspond to $\beta=2$ and $\beta=-1$ respectively, whereas a full isotropic distribution (no angular dependence) is characterized by $\beta=0$. 

In a recent work, we have examined the energy and pulse duration dependence of $\beta$ \cite{JCP.147.144304} for a single ionizing pulse. In particular, we were able to assign a well defined value of $\beta$ for each individual $N^+$ channel reached by photoionization, even with an extremely short pulse of bandwidth much larger than the rotational spacing. If we start from the initial state $N_E=0$, in the cw limit the \mbox{$N^+=0$} exit channel is characterized by a peaked anisotropy with $\beta=2$, whereas a quasi-isotropic \mbox{$\beta=1/5$} angular distribution is reached for $N^+=2$\,\cite{JCP.147.144304}.

The present work is devoted to a double-pulse control of the interference scenario between the ionization pathways referring to the excitation scheme illustrated in Fig.~\ref{fig:model}. The observable is the energy and time delay dependent asymmetry parameter $\beta$. The results are once again interpreted in terms of the simplified model depicted in Eq.\,(\ref{eq:pcos17b}), where the proportionality factor $|\mathcal{A}_{N^+}(\varepsilon)|^2$ is now also considered as $\theta$-dependent.

%%%%%%%%%%%%%%%%%%%%%%%%%%%%%%%%%%%%%%%%%%%%%%%%%%%%%%
\subsubsection{Analysis for a fixed time delay $\tau$}
%%%%%%%%%%%%%%%%%%%%%%%%%%%%%%%%%%%%%%%%%%%%%%%%%%%%%%

For the initial state $N_E=0$, a complete interpretation for the asymmetry parameter $\beta$ is provided in terms of contributions of the two ionization channels $N^+=0$ and $2$. This is achieved starting with a single pulse excitation (or equivalently, $\tau=0$) leading to the photoelectron spectrum illustrated in Fig.\,\ref{fig:fullspectra} at the energies $\varepsilon_0=4671.6$\,cm$^{-1}$ for $N^+=0$ and $\varepsilon_2=4668.6$\,cm$^{-1}$ for the $N^+=2$ ionization channel. The asymmetry parameter $\beta$ calculated from the full time evolution model (i) amounts to be $\beta(\varepsilon_0) \simeq \beta(\varepsilon_2) \simeq 1.69$. We interpret this as a mixed contribution of both ionization channels $N^+=0$ and $N^+=2$ at the two very close energies $\varepsilon_0$ and $\varepsilon_2$. A second calculation is then conducted for an arbitrarily chosen  time delay, say $\tau=7$\,ps for instance. As depicted by the dashed green lines of Fig.\,\ref{fig:betacontrol} we get $\beta(\varepsilon_0) \simeq 1.93$ for $N^+=0$ and $\beta(\varepsilon_2) \simeq 1.09$ for $N^+=2$, a result closer to the respective values $2$ and $1/5$ expected in the cw limit when compared to the identical values $\beta(\varepsilon_0) \simeq \beta(\varepsilon_2) \simeq 1.69$ obtained with a single pulse. We rationalize this finding by enhanced contributions of channel $N^+=0$ at the energy $\varepsilon_0$ and of channel $N^+=2$ at the energy $\varepsilon_2$ in an interference mechanism among the two ionization pathways created by the interaction with two time-delayed ionization pulses.

%%%%%%%%%%%%%%%%%%%%%%%%%%%%%%%%
\subsubsection{Control strategy}
%%%%%%%%%%%%%%%%%%%%%%%%%%%%%%%%

\begin{figure}[t]
\centering
\includegraphics[width = 0.99\columnwidth]{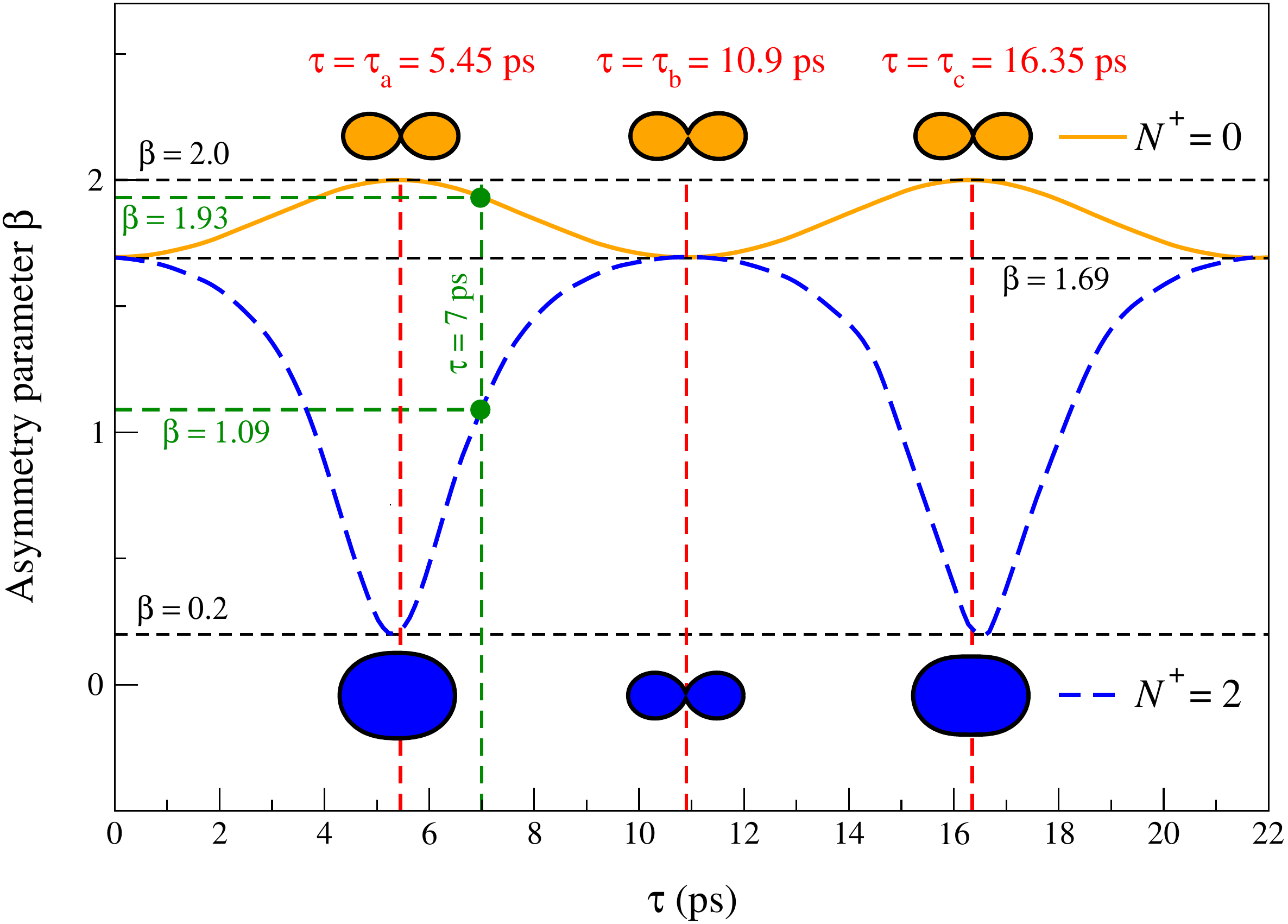}
\caption{(Color online) Asymmetry parameter $\beta$ and associated angular distributions as a function of the inter-pulse time delay $\tau$ for the initial state $N_E=0$. The middle row displays the asymmetry parameter as a function of the time delay, as evaluated at the specific energies $\varepsilon_0$ of the $N^+=0$ (orange solid line) and $\varepsilon_2$ of the $N^+=2$ (dashed blue line) exit channels. Polar representations of the photoelectron angular distributions for time delays corresponding to constructive and destructive interferences for three optimal values of $\tau$ are given in the upper and lower rows, for $N^+=0$ and $2$, respectively. The values of $\beta$ obtained for the specific delay time $\tau=7$\,ps are highlighted by additional dashed green lines.}
\label{fig:betacontrol}
\end{figure}

This claim is now supported by a control strategy aiming at a complete constructive \emph{vs.} destructive interference scheme targeting the channels $N^+=0$ and $N^+=2$, through an appropriate choice of the time delay. The choice for an optimal $\tau$ is done through the interference model (ii). The results are gathered in Fig.~\ref{fig:betacontrol}. The solid orange and dashed blue lines correspond to the variations of $\beta$ as a function of $\tau$ calculated at the energies $\varepsilon_0$ and $\varepsilon_2$ associated with the channels $N^+=0$ and $N^+=2$.

Three inter-pulse time delays $\tau_a=5.45$\,ps, $\tau_b=10.9$\,ps and $\tau_c=16.35$\,ps present a specific interest for the interference control strategy and are retained in the upper (for $N^+=0$) and lower (for $N^+=2$) rows which display polar-angle plots of the corresponding photoelectron angular distributions in each individual channel. These pulse delays were chosen because they correspond to $\tau_a = \tau_0^*(\varepsilon_2) = \tau_2^*(\varepsilon_0) = \pi\hbar/(6B)$, $\tau_b = 2\tau_1 = \pi\hbar/(3B)$ and $\tau_3 = 3\tau_1 = \pi\hbar/(2B)$. $\tau_a$ is therefore the first delay time associated with a destructive interference as defined in Eq.\,(\ref{eq:Tdes}) for $n=0$. Similarly, $\tau_b$ is the first non-zero delay time associated with a constructive interference as defined in Eq.\,(\ref{eq:Tcons}) (for $n=1$). Finally, $\tau_c$ is the second delay time associated with a destructive interference.

It is worthwhile emphasizing that a photoelectron probability at a given energy is build up from the incoherent contributions of both the $N^+=0$ and the $N^+=2$ channels with appropriate weighting coefficients. According to Eq.\,(\ref{eq:pcos_des}), a time delay associated to a destructive interference such as $\tau_a$ and $\tau_c$ imposes that, at the energy $\varepsilon_0$, the weighting coefficient of the $N^+=2$ component is zero, whereas the one of the other component $N^+=0$ is one. For the same time delay, at the energy $\varepsilon_2$ the weighting coefficient of the $N^+=0$ component is zero, whereas the one of $N^+=2$ is one. This clear separation of the respective contributions of the $N^+=0$ and $N^+=2$ channels at their respective energies $\varepsilon_0$ and $\varepsilon_2$ happens for $\tau=\tau_a$ and for $\tau=\tau_b$ due to a destructive interference induced by a controlled double pulse excitation. On the contrary, the interference is constructive at the intermediate delay time $\tau_b$, yielding a mixed contribution of the two components $N^+=0$ and $2$ at both energies. We see in Fig.\,\ref{fig:betacontrol} that this particular time delay $\tau_b$, characterized by $\beta=1.69$ in both channels, leads to two identical angular distributions at the energies $\varepsilon_0$ and $\varepsilon_2$, with a marked anisotropy similar to the one of the dominant $N^+=0$ ionization channel in the cw limit. The only noticeable difference is that at $\theta=\pm\pi/2$ (perpendicular to the polarisation axis), the two loops of the polar plots are topologically not strictly closed. Returning back to the time delays $\tau_a$ and $\tau_c$, the induced double pulse interference now gives rise at the energy $\varepsilon_0$ (upper row) to a $\beta$ parameter strictly equal to $2$ and therefore to the anisotropic polar plot which is the one expected for $N^+=0$ using cw light. Reciprocally, the resulting double pulse interference at the energy $\varepsilon_2$ (lower row) goes with a $\beta$ parameter equal to $1/5$ and with a quasi-isotropic polar plot, identical to the one expected for the $N^+=2$ exit channel in the cw limit. The destructive interference predicted for the time delays $\tau_a$ and $\tau_c$ is therefore clearly observed. This interference effect leads, for the specific time delays $\tau=(2n+1)\,\tau_a$, to a clear separation of the different exit channels at their associated energies and therefore to an angular distribution which is the one expected in the cw limit even with extremely short pulse.

\begin{figure}[t]
\centering
\includegraphics[width = 0.99\columnwidth]{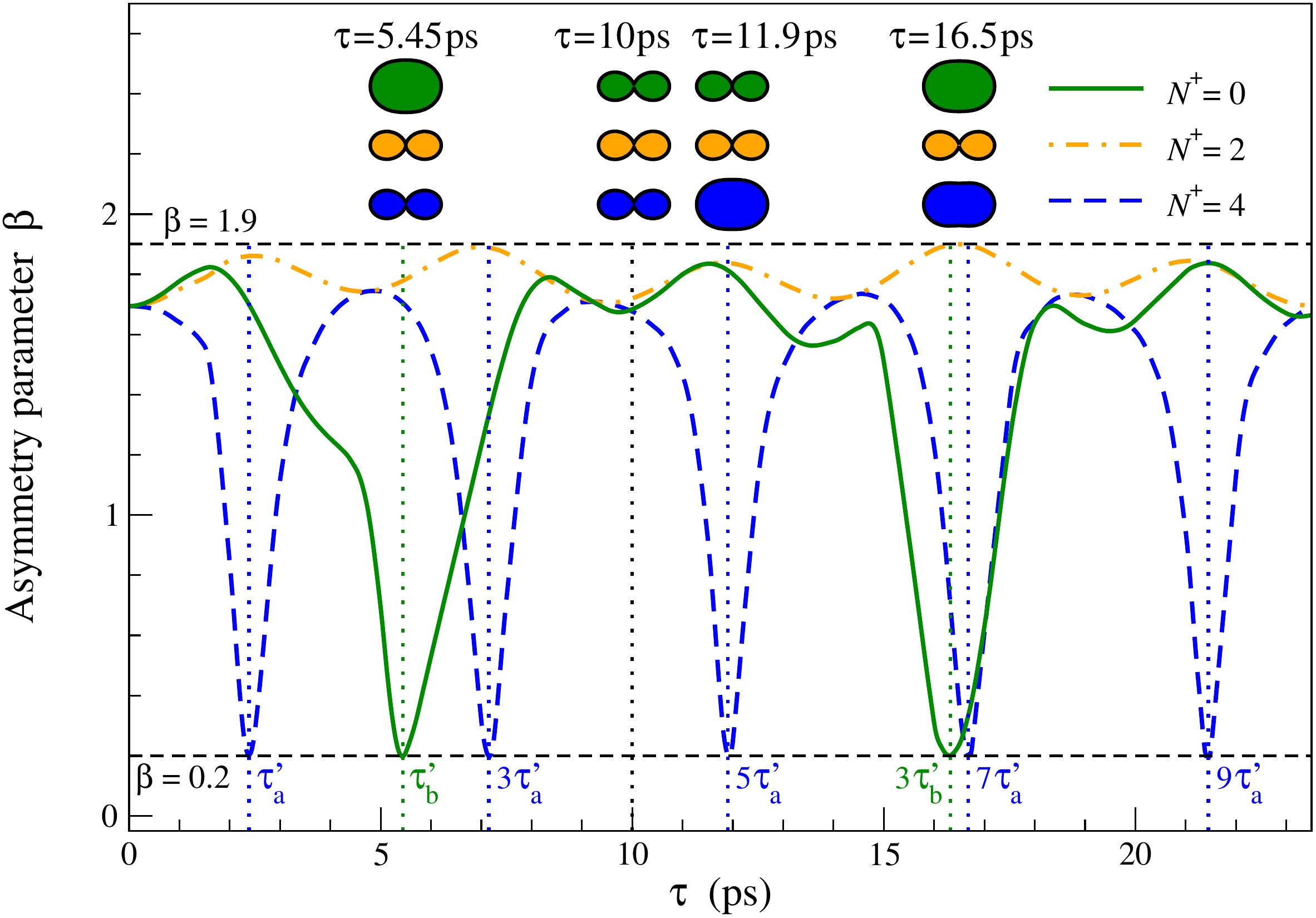}
\caption{(Color online) Asymmetry parameter $\beta$ and  associated  angular  distributions as a function of the inter-pulse time delay $\tau$ for the initial state $N_E=2$. The  middle row displays the asymmetry parameter as a function of the time delay, as evaluated at the specific energies $\varepsilon_0$ of the $N^+=0$ (green solid line), $\varepsilon_2$ of the $N^+=2$ (orange dash-dotted line) and $\varepsilon_4$ of the $N^+=4$ (blue dashed line) exit channels. Polar representations of the photoelectron angular distributions associated with these three different energies are given in the upper row at four specific delays $\tau = 5.45$, $10$, $11.9$ and $16.5$\,ps using the same color code. The time delays $(2n+1)\tau'_a$ and $(2n+1)\tau'_b$ where a destructive interference is predicted are indicated by blue and green vertical dotted lines (see text for details).}
\label{fig:betacontrol3}
\end{figure}

%%%%%%%%%%%%%%%%%%%%%%%%%%%%%%%%%%%%%%%%%%%%%%%%%%%%
\subsubsection{Application to another initial state}
%%%%%%%%%%%%%%%%%%%%%%%%%%%%%%%%%%%%%%%%%%%%%%%%%%%%

It is of course important to check the physical relevance of this double pulse control strategy through an appropriate choice of the time delay $\tau$, by exploring its application to the ionization process starting from an initial rotational state different from $N_E=0$. In this sub-section we are therefore exploring the case $N_E=2$. A single photon absorption followed by a $p$-type electron ejection now leads to the creation of three ion rotational states $N^+=0$, $2$, $4$, energetically distant by $6B$ for the pair $\{0,2\}$, $14B$ for the pair $\{2,4\}$ and $20B$ for the pair $\{0,4\}$. Ultra-short large bandwidth pulses are such that all these channels can be mixed during the ionization process. Once again referring to the interference model (ii) and to the fact that the expected oscillation patterns associated with each exit channel $N^+$ should be added incoherently, it could be expected that the regular succession of constructive \emph{vs.} destructive interferences observed in Fig.\,\ref{fig:betacontrol} for the simpler case $N_E=0$ could be washed out in the present and more complex case with three different allowed exit channels. However, it is important to remember that the anisotropy of Li$_2$ is relatively weak in the electronic state E$(^1\Sigma_g^+)$ and as a consequence the exit channel $N^+=2$ is favored for the initial state $N_E=2$. Out of the three pairs of ion rotational states mentioned above, the two that will give rise to a significant interference effect are therefore the pairs $\{0,2\}$ and $\{2,4\}$. Their energy splittings are $6B$ and $14B$, respectively. The time delays associated with the first destructive interference events as defined in Eq.\,(\ref{eq:Tdes}) are therefore
\begin{equation}
\tau'_a = \tau^*_2(\varepsilon_0)
        = \frac{\pi\hbar}{\varepsilon_4-\varepsilon_2}
        \simeq \frac{\pi\hbar}{14B}
        \simeq 2.38\,\mathrm{ps}
\label{eq:tap}
\end{equation}
for the pair $\{2,4\}$ and
\begin{equation}
\tau'_b = \tau^*_2(\varepsilon_4)
        = \frac{\pi\hbar}{\varepsilon_2-\varepsilon_0}
        \simeq \frac{\pi\hbar}{6B}
        \simeq 5.45\,\mathrm{ps}
\label{eq:tbp}
\end{equation}
for the pair $\{0,2\}$. The first seven destructive interference events are therefore expected for the specific time delays $\tau'_a$, $\tau'_b$, $3\tau'_a$, $5\tau'_a$, $3\tau'_b$, $7\tau'_a$ and $9\tau'_a$ corresponding to the values $\tau = 2.38$\,ps, $5.45$\,ps, $7.15$\,ps, $11.92$\,ps, $16.35$\,ps, $16.68$\,ps and $21.45$\,ps.

\begin{figure}[t!]
\centering
\includegraphics[width = 0.99\columnwidth]{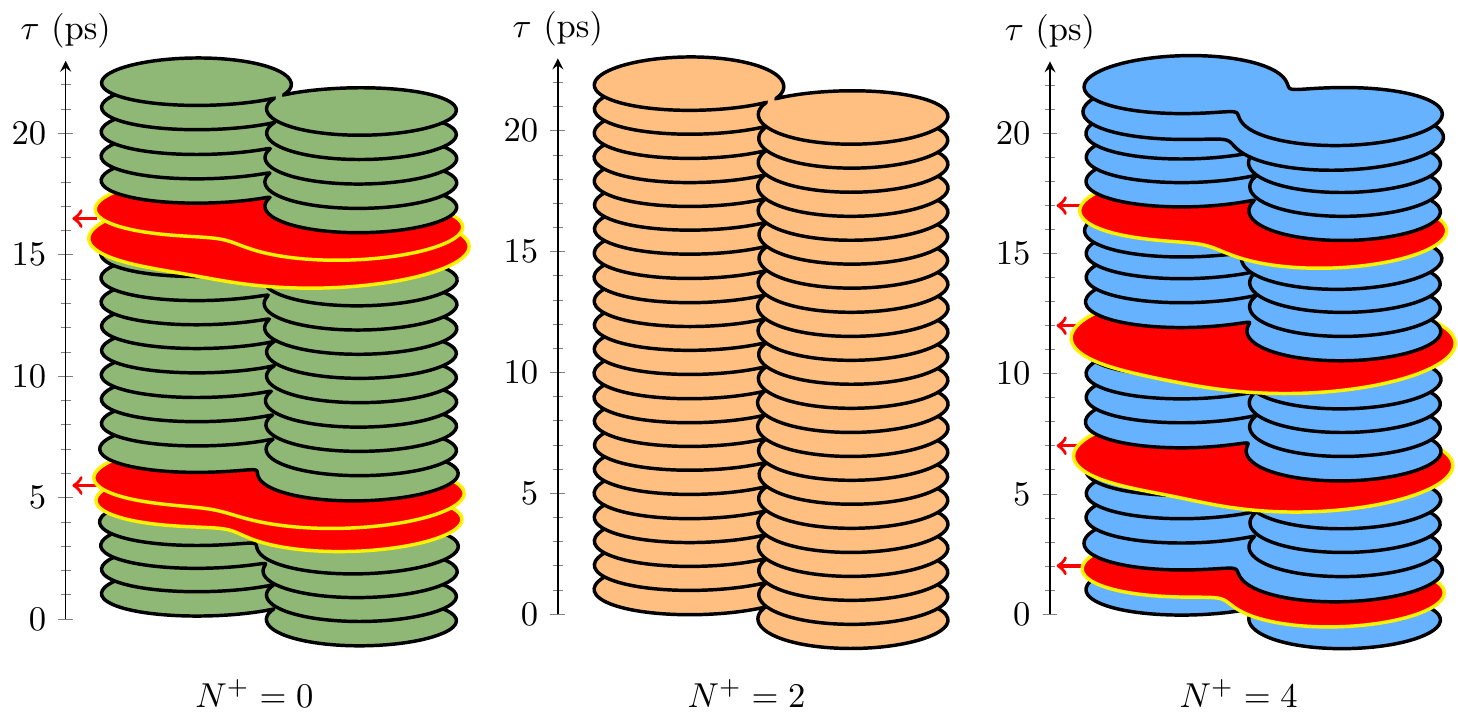}
\caption{(Color online) Polar plots of the photoelectron angular distributions as a function of the inter-pulse time delay $\tau$ for an initial state $N_E=2$ at the energies $\varepsilon_0$ (left side in green), $\varepsilon_2$ (middle column in orange) and $\varepsilon_4$ (right side in blue) corresponding to the three possible exit channels $N^+=0$, $2$ and $4$. The polar plots are calculated with increasing values of $\tau$ (every 1\,ps) and are vertically packed, as indicated by the three vertical axes. The angular distributions in channels $N^+=0$ and $N^+=4$ affected by a destructed interference (see the corresponding behaviors of $\beta$ in Fig.\,\ref{fig:betacontrol3}) are emphasized as red-filled surfaces. See text for details.
}
\label{fig:angldis}
\end{figure}

The results of the numerical simulation are displayed in Fig.\,\ref{fig:betacontrol3} where the variations of $\beta$, calculated at the energies $\varepsilon_0$ (green solid line), $\varepsilon_2$ (orange dash-dotted line) and $\varepsilon_4$ (blue dashed line) corresponding to the three possible exit channels, are plotted against $\tau$. The first seven destructive interference events discussed above are highlighted by blue (for the pair $\{2,4\}$) and green (for the pair $\{0,2\}$) vertical dotted lines. They all correspond to a minimum of $\beta$, very close to $1/5$, as expected for the $N^+=0$ and $N^+=4$ ionization channels. The corresponding photoelectron angular distribution are shown as polar-angle plots in the upper row of Fig.\,\ref{fig:betacontrol3} for each individual channel. They confirm the very strong influence of the interference effect on the secondary ionization channels $N^+=0$ and $N^+=4$. An interesting circumstance concerns the critical time delays which are common odd multiples of the periods $\tau'_a$ and $\tau'_b$, such that the values of the asymmetry parameter $\beta$ on both channels $N^+=0$ and $4$ are simultaneously minimum. This mathematically amounts  to search integer values $(n_a, n_b)$ such that $(2n_a+1)\tau'_a=(2n_b+1)\tau'_b$. From Eqs.~(\ref{eq:tap}) and (\ref{eq:tbp}) we see that a possible couple of integers is found for $n_a=3$ and $n_b=1$, leading to a critical time delay close to $7\tau'_a \simeq 3\tau'_b \simeq 16.5$\,ps. The polar representations of the angular distributions calculated for this particular time delay are shown in the upper row of Fig.\,\ref{fig:betacontrol3}. They confirm the simultaneous destructive interference predicted by the reduced model (ii) in the two secondary ionization channels. We also see from Fig.\,\ref{fig:betacontrol3} that for a randomly chosen delay time such as $\tau=10$\,ps all accessible ionization channels are mixed and their associated angular distributions are characterized by an anisotropic shape similar to the one of the dominant channel $N^+=2$. Fig.~\ref{fig:angldis} illustrates and summarizes the kind of control which is exerted in the three exit channels $N^+=0$, $2$ and $4$ by varying the control parameter $\tau$. This is given in terms of polar views of photoelectron angular distributions vertically packed with increasing $\tau$, every 1\,ps. The conclusion is that, except in the dominant channel $N^+=2$, where the distributions are basically identical for all values of $\tau$, specific values of the inter-pulse delay time can be found (see Fig.~\ref{fig:betacontrol3}), for which marked quasi-isotropic behaviors are obtained in the secondary channels $N^+=0$ and $N^+=4$. These particular delay times are emphasized in Fig.~\ref{fig:angldis} as red-filled surfaces and correspond to the specific delay times $\tau'_a$, $\tau'_b$, $3\tau'_a$, $5\tau'_a$, $3\tau'_b$, and $7\tau'_a$ mentioned previously.

Finally, this analysis confirms the validity of the simple model (ii) for the interpretation and for the prediction of the influence of the interferences induced by a double-pulse ionization of the molecule on the photoelectron angular distributions in the ultra-short regime.

%%%%%%%%%%%%%%%%%%%%%%%%%%%%%%%%%%%%%%
\subsubsection{Practical implications}
%%%%%%%%%%%%%%%%%%%%%%%%%%%%%%%%%%%%%%

For the purpose of practical implementation, four regimes are particularly interesting. These are depicted in Fig.~\ref{fig:betacontrol3} with vertical blue and green dotted lines on top of which polar diagrams of photoelectron angular distributions are displayed. Butterfly-like diagrams correspond to anisotropic distributions along the polarization axis ($\beta$ close to $2$), while oblate disks correspond to quasi-isotropic distributions with $\beta$ close to $1/5$. The variety of diagrams illustrate the richness of the monitoring that can be achieved by changing the time delay between the pump and the probe pulses.

The first regime appears for $\tau \simeq 5.45$\,ps, where the fast electrons associated with the $N^+=0$ ionization channel show a dramatic change in $\beta$ downwards a quasi-isotropic distribution. In this regime, it will be possible to selectively detect these fast electrons in the plane perpendicular to the polarization axis since the probability of electron emission from the other channels is then minimum. The reciprocal behaviour happens for the slow electrons associated with the channel $N^+=4$ at $\tau \simeq 11.9$ ps. Here again, a sharp discrimination of this channel from the others is possible thanks to the net dip-like evolution of $\beta$, whereas the two other channels produce peaked electron beam distributions along the polarization axis. A third regime occurs around $\tau \simeq 10$\,ps. Here, all the photoionized electrons follow the same quasi-linear path along the polarization direction. Finally, the regime that appears around $\tau \simeq 16.5$\,ps, is particularly interesting. Indeed, it corresponds to a synchronisation of the two stroboscopic interference effects where both fast and slow electrons follow quasi-isotropic distributions, as illustrated with the two oblate disks on the top of the figure at this particular value of the time delay. The same effect can be seen in Fig.\,\ref{fig:angldis}, where red flat disks emerge from the green $(N^+=0)$ and the blue $(N^+=4)$ vertical stacks in the region $16\,\mathrm{ps} \leqslant \tau \leqslant 17\,\mathrm{ps}$. Moreover, these two ionization channels are in principle dominated by the main channel $N^+=2$, whose probability is much higher, but thanks to the aforementioned synchronisation there is a chance to detect selectively the minority electron channels $N^+=0$ and $4$ in the perpendicular direction, where both contributions are added so that the signal will be enhanced and therefore measured more efficiently. This kind of experiment may also be performed using refined time-of-flight techniques, where photoionization channels can be detected separately, by collecting electrons and Li$_2^+$ residual ions in coincidence \cite{Dowek_2012}.

%%%%%%%%%%%%%%%%%%%%%%%%%%%%%%%%%%%%%%%%%
\section{Conclusions and perspectives}
\label{sec:conclusion}
%%%%%%%%%%%%%%%%%%%%%%%%%%%%%%%%%%%%%%%%%

A detailed understanding and control of photoionization processes can be conducted based on measurements of photoelectron spectra involving both electron kinetic energy and angular distributions as observables. This can be achieved by using a train of two ultra-short pulses separated by a time delay $\tau$ taken as a control knob. Considered individually, due to their large bandwidth, each of these pulses leads to a single broad peak in the resulting photoelectron spectrum, which comprises not yet resolved contributions of different ion rotational channels. Phases accumulated during the inter-pulse time delay can however produce interference behaviors among these exit channels with well resolved and controllable oscillating patterns in photoelectron spectra. Such a control is based on constructive or destructive interference mechanisms among the ion rotational channels, achieved in a robust way through a single parameter $\tau$, arguing thus in favor of 
experimental feasibility. Moreover, molecular asymmetry parameters $\beta$ associated with individual ion rotational channels can be extracted and controlled, to reach desired angular distributions from quasi-isotropic to highly anisotropic. It is worthwhile noting that such a situation is unexpected for a single ultra-short pulse, with the challenge to address an asymmetry parameter $\beta$ to individual ion rotational channels.

Li$_2$ is a theoretically and experimentally well documented illustrative example that we have chosen for this study. However, the interference model and control strategy developed in this work are generic enough to be transposable to other diatomic molecules and offer the possibility of a detailed understanding of the ionization process in terms of coherent dynamics among rotational exit channels. As future perspectives, the role of interfering ion vibrational channels could be considered with similar strategies, eventually referring to additional ultra-short pulses in the train with different time delays offering additional flexibility to the control scheme. Moreover the model, which accounts for both electronic and nuclear degrees of freedom, could be extended to mechanisms underlying dissociative ionization and control strategies, with coincidence measurements as observables.

%%%%%%%%%%%%%%%%%%%%%%%%%%
\section*{Acknowledgment}
%%%%%%%%%%%%%%%%%%%%%%%%%%
This work is supported by the ``Investissements d{'}Avenir'' Program of the Labex PALM of the University Paris-Saclay under grant number ANR-10-LABX-0039-PALM. RC acknowledges financial support from ISMO, for a stay during which part of this work was achieved. We also acknowledge the use of the computing cluster MesoLum/GMPCS of the LUMAT research federation (FR 2764 CNRS).

%%%%%%%%%%%%%%%%%%%%%%%%%%%%
%
%%%%%%%%%%%%%%%%%%%%%%

\end{document}